 \documentstyle[12pt]{article}
\newskip\humongous \humongous=0pt plus 1000pt minus 1000pt
\def\caja{\mathsurround=0pt}

\def\eqalign#1{\,\vcenter{\openup1\jot \caja
        \ialign{\strut \hfil$\displaystyle{##}$&$
        \displaystyle{{}##}$\hfil\crcr#1\crcr}}\,}
\newif\ifdtup

\def\be{\begin{equation}}
\def\ee{\end{equation}}
\def\ba{\begin{eqnarray}}
\def\ea{\end{eqnarray}}

\def\e{\epsilon}
\begin{document}
\renewcommand{\theequation}{\thesection.\arabic{equation}}
\newcommand{\beq}{\begin{equation}}
\newcommand{\eeq}[1]{\label{#1}\end{equation}}
\newcommand{\ber}{\begin{eqnarray}}
\newcommand{\eer}[1]{\label{#1}\end{eqnarray}}
\begin{titlepage}
\begin{center}

\hfill NSF-ITP/95-144\\
\hfill CPTH-S388.1195\\
\hfill hep-th/9511043\\

\vskip .4in

{\large \bf  D-BRANE DYNAMICS}
\vskip .4in

{\bf C. Bachas}
\vskip .15in

{
\em Institute for Theoretical Physics, University of California,
  Santa Barbara, CA 93106, USA \\
   and \\
  Centre de Physique Th\'eorique, Ecole Polytechnique,
 91128 Palaiseau, FRANCE\\
email: bachas@orphee.polytechnique.fr
}

\vskip .15in

\end{center}

\vskip .5in

\begin{center} {\bf ABSTRACT }\\
\end{center}

{
I calculate
the semiclassical
 phase shift ($\delta$),
as function of impact parameter
($b$)
and velocity ($v$),
 when one D-brane moves
  past another. From its low-velocity expansion
I show that,
for torroidal compactifications, the moduli space
of two identical D-branes stays flat to all orders in $\alpha^\prime$.
For K3 compactifications, the calculation of the D-brane
moduli-space metric can be
mapped to a dual gauge-coupling renormalization problem.
In the ultrarelativistic regime, the absorptive part of the
phase shift grows as if the
D-branes were  black disks of area
$\sim \alpha^\prime ln{1\over 1-v^2}$.
The scattering
of large fundamental strings shares all the above qualitative
features.
A side remark concerns the intriguing duality between limiting
electric fields and the speed of light.
 }

\begin{quotation}\noindent
\end{quotation}
\vskip 0.7cm
November 1995\\
\end{titlepage}
\vfill
\eject
\def\baselinestretch{1.2}
\baselineskip 16 pt
\noindent
 \setcounter{equation}{0}

{\it Introduction.}
D(irichlet)-branes \cite{Joe1,Joe2,Leigh,Joe3}
are dynamical extended defects, described by the fact that
open strings have their end-points stuck on them.
In an important development Polchinski
has recently shown \cite{Joe1} that D-branes carry
unit charge
under the Ramond-Ramond gauge fields of closed type-II string
theory, so that together with their bound states \cite{Witt}
they could provide the excitations required by
various forms of string duality \cite{Hull}.
This bold conjecture makes such excitations amenable to
study with the techniques of conformal field theory.
Here I want to use such techniques to study certain aspects
of the dynamics of D-branes.
Of great help in the discussion will be the
(perturbative) duality between
brane motion and electromagnetic open-string backgrounds:
obscure statements in one language become
transparent in the other, and vice versa.
 I will therefore start by explaining briefly this duality.

\vskip 0.3cm

{\it D-brane motion and electromagnetic backgrounds.}
Consider  a 0-brane or ``point particle'' describing
some trajectory $Y^j(X^0)$ where $j$ is as usual
a spatial index. The appropriate
world-sheet action on the disk reads
$$
S = {1\over 4\pi\alpha^\prime}  \int d^2 z \
  \partial^a X^\mu \partial_a X_\mu +  {1\over 2\pi\alpha^\prime}
\int d\tau\ Y^j(X^0) \partial_{\sigma} X_j  \ , \eqno(1)
$$
where $z=\tau+i\sigma$, the bulk integral is over the
upper half plane, and the coordinates $X^0$, $X^j$ of the
string obey Neumann, respectively fixed
Dirichlet  conditions.
In writing this action we make use of the
fact that the brane coordinates couple to the boundary
vertex operator $\partial_{\sigma} X^j/(2\pi\alpha^\prime)$
 \cite{Joe2,Leigh}.
 The $\beta$-function equations for this coupling can thus be
interpreted as the classical equations of motion of the 0-brane.
Rather than do the calculation, let us map this into an
identical mathematical problem:
$$
{\tilde S} = {1\over 4\pi\alpha^\prime} \int d^2 z \
  \partial^a X^\mu \partial_a X_\mu +
i e\int d\tau\  A^j(X^0) \partial_{\tau} X_j   \ , \eqno(2)
$$
where all string coordinates obey now Neumann conditions.
This is the action on the disk in the presence
of a time-varying,
constant in space electric field $E^j=\partial_0 A^j$,
coupling to a boundary that carries charge $e$.
To see that the two problems are identical, notice that
on the boundary
$$\eqalign{
 < \partial_{\sigma} X^j(\tau) \partial_{\sigma}& X^k(\tau^\prime)>
\vert_{Dirichlet}=\cr
= - & < \partial_{\tau} X^j(\tau) \partial_{\tau} X^k(\tau^\prime)>
\vert_{Neumann}
=  {2\alpha^\prime\ \delta^{jk} \over (\tau-\tau^\prime)^2} \ ,
\cr}  \eqno(3)
$$
so that modulo zero modes
the  loop expansions of (1) and (2)
are the same.
Now the dynamics
of a slowly-varying electric field is known to be governed by
the Born-Infeld action \cite{call}.
Under the exchange $2\pi\alpha^\prime e E^j \leftrightarrow
v^j\equiv \partial_0 Y^j$ this
becomes the action for a
relativistic point particle:
$$ {\cal L}_{BI} \propto  \sqrt{1-(2\pi\alpha^\prime e{\bf E})^2}\
\ \leftrightarrow\ \
 {\cal L}_{particle} \propto  \sqrt{1-v^j v_j} \ .
\eqno(4)
$$
What we see here is a
 manifestation, noticed previously by Leigh \cite{Leigh},
 of the
  well-known perturbative duality
between Neumann and Dirichlet conditions
 \cite{Joe2,od}.
Though technically trivial, the  above identification
is all the same
startling: the Born-Infeld action
is the result of resummation of all, stringy in nature,
$\alpha^\prime$ corrections, and implies
among other things the existence of
a limiting electric field $E_{crit} = (2\pi\alpha^\prime e)^{-1}$
\cite{call,bp}.
In the dual picture on the other hand
all this is just
a consequence of the laws of
relativistic particle mechanics for the 0-brane,
 the limiting velocity being
simply the speed of light!
Put differently, the fact that the
solitons of type-II string theory should behave as
 relativistic
particles, fixes uniquely the form of the
(abelian) Born-Infeld action.

The Regge slope $\alpha^\prime$ does not in fact completely
disappear from the
0-brane dynamics. Indeed, derivatives of the electric field
  modify
the Born-Infeld action \cite{Arc}
\footnote{ It is nevertheless a curious
coincidence that
the leading two-derivative   terms, $(\partial F)^2 F^n$,
 are  absent
in the supersymmetric case \cite{Arc}.},
 so that by duality the acceleration measured
in units of $(\alpha^\prime)^{-1/2}$ should also enter
into the full 0-brane action.
This looks at first sight paradoxical: aren't  straight
world lines  after all the only allowed motions of a point
particle?
The resolution of the puzzle has to be that
$Y^j(X^0)$ loses its meaning as a point-particle
trajectory when one looks at it at scales
 shorter than $\sqrt{\alpha^\prime}$.
I will come back to this point in  the very end.

The above discussion can be extended easily to the
case where some spatial coordinates,
 $X^1, .., X^p$, enter on the same footing as $X^0$.
{}From the gauge-field point of view this corresponds to having
both electric and magnetic backgrounds.
In the dual language,  on the other hand,
  $Y^{M}(X^{\alpha})$
with $\alpha=0,1..,p$ and $M=p+1,..,d$
 now describe  the
transverse motion of a p-brane.
After some straightforward
  matrix algebra the Born-Infeld action
for the above backgrounds takes the
form,
$$ {\cal L}_{BI} \propto \sqrt{-det(\eta_{\mu\nu} +
2\pi\alpha^\prime e F_{\mu\nu})}
\ \leftrightarrow
{\cal L}_{p-brane} \propto \sqrt{-det( \eta_{\alpha\beta}
+ \partial_\alpha Y^M
\partial_\beta Y_M) }\ . \eqno(5)
$$
 Not suprisingly the right-hand side
is   the Nambu-Gotto action,
in the   gauge in which the first $p+1$ (longitudinal)
space-time coordinates
 are used to
 parametrize the world history
of the brane.
The two sides of eq. (5) can be combined more
generally into the Dirac-Born-Infeld action \cite{Leigh},
which describes electromagnetic fields on a fluctuating membrane.
Higher-order terms would again modify
  this action at string scales.
It is also interesting to comment on
what happens when some of the
coordinates are compactified. As one moves around a compact
longitudinal coordinate, the total  displacement in a
  compact transverse dimension must be an integer multiple of
the period.
 This is a classical-geometric analog
 of  Dirac's quantization
condition, with   the role of
magnetic charge being played by the transverse winding number.
\vskip 0.3cm

{\it Calculation of the phase shift}.
Uniform motion and orientation of a single
D-brane cannot have
any non-trivial consequences, since it depends on the choice
of an inertial frame.
Invariant meaning can however be attached
to the relative motion and orientation of two branes,
 which can be sensed by open strings that stretch out between them.
For instance  relative disalignement  breaks
space-time supersymmetry exactly like a magnetic field
does in the
dual case \cite{me}.
Here
 I want to concentrate, instead, on relative motion of
two parallel branes, and calculate the phase shift
 of their forward scattering amplitude.
The boundary conditions for an open string with ends attached
to two parallel moving p-branes read:
$$ \partial_{\sigma} X^{1,...p} = X^{p+1,..,d-2} = 0 \ \ \ \ \
{\rm at} \ \ \sigma=0,\pi  \eqno(6a)
$$
and
$$\eqalign{
 X^{d-1}&=X^{d}-v_1 X^0 = \partial_\sigma (v_1 X^d-X^0) = 0 \ \
\ \ \
{\rm at} \ \ \sigma=0  \cr
\ \ \ X^{d-1}-b &= X^{d}-v_2 X^0 = \partial_\sigma (v_2 X^d-X^0) = 0 \ \
\ \ \
{\rm at} \ \ \sigma=\pi  \ . \cr
}\eqno(6b)
$$
Here
$v_1$ and $v_2$ are the brane velocities in the transverse $X^d$
direction, and $b$ is the impact parameter. Notice that
the last conditions (6b) are a
consequence of the world-sheet
boundary equations of motion. The minus
sign in them is due to the Minkowski signature and plays a
 crucial role in
 what follows. Now
the coordinates $X^{1,..,d-1}$
  are either pure Neumann or pure Dirichlet
  and have conventional mode expansions.
The mode expansions of the  remaining two coordinates
can be worked out easily with the result:
$$\eqalign{\ \
X^0\pm X^d = i\sqrt{\alpha^\prime} &\sqrt{1\pm v_1 \over 1\mp v_1}
\times \cr \times
 & \sum_{n=-\infty}^{\infty}
\Bigl\{ {a_n \over n+i\epsilon}
 \e^{-i(n+ i\epsilon)(\tau\pm\sigma)} +
{{\tilde a}_n \over n-i\epsilon}
 \e^{-i(n- i\epsilon)(\tau\mp\sigma)}
\Bigr\}\ , \cr}
\eqno(7)
$$
where
$$ \pi \epsilon \equiv  {\rm arctanh}(v)
=   {\rm arctanh}(v_2)-  {\rm arctanh}(v_1) \ .
\eqno(8)
$$

\noindent
We will assume from now on that $v_2>v_1$ so that
$\epsilon$ is positive.
Reality imposes the conditions
 $a_n^*=a_{-n}$ and ${\tilde a}_n^* = {\tilde a}_{-n}$,
while the canonical commutation relations imply
$$
[ a_n, {\tilde a}_m] = (n+i\epsilon) \delta_{n+m,0} \ .
\eqno(9)
$$
Finally the total world-sheet Hamiltonian reads
$$
L_0 =  {b^2\over 4\pi^2 \alpha^\prime} +
\sum_{n=1}^\infty a_n^* {\tilde a}_n +
\sum_{n=0}^\infty   {\tilde a}_n^* a_n
+{i\epsilon (1-i\epsilon)\over 2} + {\rm standard} \ ,
\eqno(10)
$$
where ``standard'' are   the
 contributions of the pure Neumann or
 Dirichlet coordinates $X^{1,. .,d-1}$,
other than the impact-parameter dependent piece which we
 explicitly exhibit.
The net effect of the brane motion
on stretched strings   can
  be now
 summarized as follows: {\it (i)} the
 frequencies of oscillation in the $(X^0,X^d)$
hyperplane are shifted by $\pm i\epsilon$, and
  {\it (ii)} there is an overall velocity-dependent
 energy subtraction. Technically it is as if the
stretched strings
belong to  a twisted sector
of an orbifold  with imaginary twist angle.

 All this is again analogous to the spectrum of
free open strings in a constant electric-field
background
\cite{call,bp}.
Different velocities correspond  to different end-point
charges, while the expression (8) for the twist parameter,
which  has no obvious interpretation
in the electric-field case, is here recognized
as the relativistic
composition of velocities of the two branes.
Indeed, as
  dictated by Lorentz invariance, the spectrum only
depends on the velocity $v$
of one brane in the rest frame of the other.
The annulus diagram, which gave
 the induced one-loop
Lagrangian in the electric-field case, corresponds now to the
phase shift for the forward scattering of two D-branes
\footnote{
This is true  for D-branes
in type-II theory, since
 type-I D-branes may also  exchange an  open string corresponding to
 the disk topology.}
{}.
The extension of the analysis to superstrings,
  as well as the calculation of the annulus diagram
are both
straightforward, when one keeps the orbifold analogy in mind.
They have been
worked out in detail for the electric field in ref. \cite{bp},
so I will refrain from repeating the parallel steps in our case.
 The only
significant difference is the absence of zero modes for
$X^0\pm X^d$, which   correctly  accounts for
 the fact that the
  D-branes interact locally in transverse space and in
 time. The final expression for the phase shift in the
supersymmetric case reads:
\footnote{ The factor of 2 in front
takes care of the fact
that for oriented strings
 one can change the role of the two endpoints\cite{Joe1}.
As a check of the normalizations
note that in the limit of vanishing velocity,
eq. (11)
   reduces formally to the result
of the adiabatic approximation
in which the branes are treated as quasi-stationary
 at a separation
$v\tau$ in the $X^d$ direction:
$$\eqalign{
\ \ \ \ \delta \rightarrow - V^{(p)}  & \int_0^\infty {dt\over t}\
(2\pi^2t)^{-(p+1)/2}\times\cr \times
 &\sum_{\alpha=2,3,4}
   {1\over 2} e_\alpha  \Theta_\alpha^4(0\vert {it\over 2})
  \eta^{-12}({it\over 2})\
 \ \int_{-\infty}^{\infty}  d\tau  e^{-(b^2+v^2\tau^2)  t/2\pi} \ .\cr}
$$
Note also that if the branes wrap around a compact torus,
one should replace
$ (2\pi^2 t)^{p/2}$ by a discrete momentum sum in the
above expression.
}

 $$ \eqalign{
\ \  \delta(b,v)
 = -2\times { V^{(p)} \over 4\pi}  \int_0^\infty &{dt\over t}\
(2\pi^2t)^{-p/2}\  e^{-b^2 t/2\pi}\
{ \Theta_1^\prime (0\vert {it\over 2})
 \over \Theta_1 ({\epsilon t\over 2}\vert {it\over 2})}\times \cr
&\times
 \sum_{\alpha=2,3,4}
\  \ {1\over 2}
e_\alpha \Theta_\alpha ({\epsilon t\over 2}\vert {it\over 2})
 \  \Theta_\alpha^3 (0\vert {it\over 2})\  \eta^{-12}( {it\over 2})
\cr}
  \eqno(11)
$$
where $e_2=-e_3=e_4=-1$,  the volume $V^{(p)}$   of the p-branes
  should be simply  dropped in the special case $p=0$,
$\Theta_\alpha$ and $\eta$ are the usual Jacobi and Dedekind
functions,
and finally we have  set $2\alpha^\prime = 1$
{}.
Expression (11) is our basic formula. It was obtained by
treating the D-branes as classical sources, and by
neglecting higher world-sheet topologies:
both the Compton wavelength and the Schwarzschild radius
of the D-brane are thus taken to vanish.
Our result is
on the other hand exact as function of $b$, $v$ and
$\alpha^\prime$.
\vskip 0.3cm

{\it Flatness of moduli space.}
Consider the behaviour of the phase shift
in the low-velocity limit,
$\pi\epsilon \simeq v \ll 1$.
The spin-structure sum in expression (11) starts out
in this limit at
quartic order,
$$
\sum_{\alpha=2,3,4}\  e_\alpha
\Theta_\alpha (\nu\vert\tau) \Theta_\alpha^3(0\vert\tau)
 \ \sim o(\nu^4) \ .  \eqno(12)
$$
This follows from the well-known supersymmetry identity, and the
fact that the $\Theta$ functions solve
 the diffusion equation \footnote{They are also even functions
of their first argument, so that odd powers of $v$ will automatically
vanish.}
 $$ [\partial_\tau + {i\over 4\pi}\partial_\nu^2]\
\Theta_\alpha(\nu\vert\tau)
 = 0 \ . \eqno(13)
$$
 The absence of the zeroth-order term in this expansion
is due to the cancellation
of gravitational attraction and Ramond-Ramond repulsion
for static D-branes \cite{Joe1}.
The absence of the quadratic term, on the other hand,
shows that the $o(v^2)$ forces also
vanish, i.e. that identical type-II
 D-branes do not scatter at very low velocities.
Put differently,   the moduli space of two D-branes is
exactly  flat
to all orders in the $\alpha^\prime$ expansion.
This statement will evidently stay true if any number
of coordinates is torroidally compactified.
In dual language it corresponds to the well-known fact that
for maximally-supersymmetric theories the Maxwell term
($\sim F^2$) is not renormalized. Based on this duality we may
in fact conjecture
 that the moduli space stays flat when
higher string-loop corrections are taken into account.
In open-string
compactifications that break
half of the space-time supersymmetries \cite{Aug},
on the other hand,
 the Maxwell term is generically  renormalized
at the one open-string-loop level.
This implies by duality
that in generic
type-II compactifications on a K3 surface,
the moduli
space of   D-branes wrapping around the entire
K3  wont be  flat, but that it
may be determined at large scales
entirely by  a single closed-string exchange. Note that
the potential infrared divergences of the open-string channel are
cutoff by the impact parameter ($b$) in expression (11).

The flatness of moduli space seems to be a consequence of maximal
supersymmetry. It
 has been established previously
 for fundamental type-II strings and for
neutral fivebranes \cite{mm,Harv}, while
 fundamental and solitonic heterotic strings
 were shown to
exhibit non-trivial scattering in the
low-velocity limit \cite{Harv}.
Our conclusions for D-branes are compatible with these results,
and are valid to all orders in the $\alpha^\prime$ expansion.

 \vskip 0.3cm

{\it Absorptive part.}
Let us turn now to the imaginary
part of the phase shift, which
arises from the zeroes of
 $\Theta_1$ at integer values of its  argument:
  $ \epsilon t/2 = k=1,2,3...$
The corresponding
 poles must all be traversed on the same side,
so as to allow a rotation of the $t$-integration axis
in the complex plane. The absorptive part can therefore
be obtained by
  summing the residues of the
integrand on these poles. The result after some
straightforward manipulations reads:

$$
{\rm Im} (\delta) = {V^{(p)}\over 2(2\pi)^p}\
 \sum_{k=1}^\infty\  {1\over k} \Bigl(
{\epsilon\over k}\Bigr)^{p\over 2}\
e^{-b^2 k/\pi\epsilon}\ \times
\Bigl\{ Z_{ferm}({ ik\over \epsilon }) - (-)^k
 Z_{bos}({ik\over \epsilon }) \Bigr\} \ ,
\eqno(14)
$$

\noindent
where $Z_{ferm}=  {1\over 2} \Theta_2^4/\eta^{12}$  and
$Z_{bos} =   {1\over 2} (\Theta_3^4-\Theta_4^4)
/\eta^{12} $
are the usual
bosonic and fermionic open-string partition functions.
This dissipation rate
is the dual counterpart of
open-string pair production in a constant
electric background \cite{bp}.
Its physical interpretation in our case is
as follows:  a pair
of open strings stretching between the two branes can
nucleate out of the vacuum and slow down or stop
 the relative motion.

Now the energy cost
for nucleating two strings must be
gained back through stretching due to the brane motion,
so we should expect this tunneling phenomenon to be
exponentially suppressed at
low velocity.
Indeed for
$\pi\epsilon \simeq v \ll 1$
only massless open-string
states contribute to the partition functions in (14)
and
the dissipation rate
reduces to
$$
{\rm Im} (\delta) \ \simeq\  \  {8 V^{(p)}\over (2\pi)^p}
 \sum_{k=1}^\infty \ {1\over k}
\Bigl(
{  v \over \pi k}\Bigr)^{p\over 2}\
e^{-b^2 k/ v }\ \ .
\eqno(15)
$$
This is  vanishingly small all the way down to
impact parameters $b  \sim \sqrt{v \alpha^\prime}$,
i.e. substantially shorter than the
fundamental string scale.
 In the ultrarelativistic
limit, on the other hand, we have
  $$\epsilon \simeq -{1\over 2\pi}
log {1-v\over 2} \simeq {1\over\pi}
log( {s\over {\cal M}_p^2}) \gg 1
\eqno(16)
 $$
with ${\cal M}_p$ the mass
of the p-brane.
The absorptive part
 is now dominated by
  the asymptotics of the
open-string partition functions
near the origin, and reads
$$ {\rm Im} (\delta)
 \simeq   {V{(p)}\over 2(2\pi)^p \sqrt{\pi}^{p-8}}\
{s\over {\cal M}_p^2}
\Bigl( log  {s\over {\cal M}_p^2} \Bigr)^{{p\over 2}-4} \times
e^{-b^2/ log( {s\over {\cal M}_p^2})}
\eqno(17)
$$
This exhibits the   characteristic behaviour of
scattering of fundamental strings
at very high energies \cite{Ven}.
The D-branes behave in this limit like black absorptive
disks of logarithmically growing area,
 $b_{cr}^2 \sim \alpha^\prime ln (s/{\cal M}_p^2)$.
As is the case for
 highly-energetic dilatons and gravitons
\cite{Kleb}, ultrafast p-branes cannot probe each other
at sub-stringy scales.
\vskip 0.3cm

{\it Outlook.}
The above results can be considered as dynamic evidence that
  D-branes behave like fundamental strings.
Several extensions of the analysis are being envisaged:
an
explicit calculation of the metric of moduli-space for
D-branes in K3 compactifications, a
  study of
  polarization dependence   through the insertion of
spin-flip operators on the boundaries of the annulus, and
  a study of the phase shift for a
  brane moving past an anti-brane
\footnote{ This was suggested  to me by J. Polchinski,
following a remark by T. Banks and L. Susskind that the force
between a static brane and anti-brane at subcritical
separation diverges.}

Do our results shed any light on Shenker's intriguing
conjecture \cite{Shenk} for
 a new dynamical scale in string theory, that
 involves the string coupling constant?
The one thing we can safely conclude is that
fast probes have no chance of detecting such a scale.
The
low-velocity scattering of torroidally-compactified
D-branes, on the
other hand, appears indeed
to be trivial at distances much shorter than the
string size. As our discussion of D-brane coordinates however
showed, it
is very questionnable whether such a
well-localized probe can be prepared as an initial state
in the first place.

\vskip 0.5cm
\noindent
{\bf Acknowledgments} \\
I am grateful to A. Tseytlin and
 G. Veneziano for useful comments,
and in particular to J. Polchinski for many illuminating
discussions.
This work was supported by the NSF under grant no. PHY94-07194, and
by EEC grants CHRX-CT93-0340 and SC1-CT92-0792.
\noindent

\vskip 0.7cm


\end{document}